\documentclass[pra,twocolumn,showpacs]{revtex4-1}%
\usepackage{dcolumn}
\usepackage{bbding}
\usepackage{natbib}
\usepackage{xcolor}
\usepackage{amsmath,amsthm,amssymb,amsfonts,mathrsfs,amsthm}
\usepackage{dsfont,yfonts,calligra}
\usepackage{graphicx,hyperref}
\usepackage{pxfonts,upgreek}
\usepackage{lmodern}
\providecommand{\U}[1]{\protect\rule{.1in}{.1in}}
\begin{document}

\title{\textbf{R{\'e}nyi generalizations of quantum information measures}}

\author{Mario Berta}
\affiliation{Institute for Quantum Information and Matter, California
Institute of Technology, Pasadena, California 91125, USA}

\author{Kaushik P. Seshadreesan}

\affiliation{Hearne Institute for Theoretical Physics, Department of Physics
and Astronomy, Louisiana State University, Baton Rouge, Louisiana 70803, USA}

\author{Mark M. Wilde}

\affiliation{Hearne Institute for Theoretical Physics, Department of Physics
and Astronomy, Louisiana State University, Baton Rouge, Louisiana 70803, USA}

\affiliation{Center for Computation and Technology, Louisiana State
University, Baton Rouge, Louisiana 70803, USA}

\newcommand{\kau}[1]{ {\color{blue}{#1}}}
\newcommand{\mario}[1]{ {\color{red}{#1}}}

\begin{abstract}
Quantum information measures such as the entropy and the mutual information find applications in physics, e.g., as correlation measures. Generalizing such measures based on the R\'enyi entropies is expected to enhance their scope in applications. We prescribe R\'{e}nyi generalizations for any quantum information measure which consists of a linear combination of von Neumann entropies with coefficients chosen from the set $\left\{-1,0,1\right\}$. As examples, we describe R\'enyi generalizations of the conditional quantum mutual information, some quantum multipartite information measures, and the topological entanglement entropy. Among these, we discuss the various properties of the R\'enyi conditional quantum mutual information and sketch some potential applications. We conjecture that the proposed R{\'e}nyi conditional quantum mutual informations are monotone increasing in the R{\'e}nyi parameter, and we have proofs of this conjecture for some special cases.
\end{abstract}

\maketitle

\section{Introduction}
\label{intro}
Quantum information theory shares large common grounds with a multitude of areas in physics. The von Neumann entropy, which is a central measure of information, for instance, is widely used under the name of entanglement entropy to study entanglement in ground states of quantum many body systems and lattice systems~\cite{VLRK_03, DHHLB_05}, relativistic quantum field theory~\cite{Calabrese_04, CC09}, and the holographic theory of black holes~\cite{Solodukhin_11, HT07}. The entanglement entropy is calculated as the entropy of the reduced state of a subsystem which consists of any bounded region in a large system, where the overall system is assumed to be in a pure state. The entanglement entropy captures the entanglement shared between the subsystem and the rest of the system across the boundary. A large body of work in the literature has focused on various questions related to the entanglement entropy, such as its scaling with respect to the size of the subsystem. The area law establishes that this scaling depends only on the area of the boundary~\cite{Bombelli_86, Srednicki_93, PEDC_05,H_07,  Eisert_10, BH_13}.

Recently, the class of $\alpha$-R{\'e}nyi entropies~\cite{Renyi_60} has been considered in the above contexts~\cite{Franchini_08, Headrick_10, ELAC_13}. The $\alpha$-R{\'e}nyi quantum entropy is a generalization of the von Neumann entropy. For a state $\rho_{A}$ on a quantum system $A$, the $\alpha$-R{\'e}nyi entropy is defined as
\begin{equation}\label{renyient}
H_{\alpha}(  A)  _{\rho}\equiv\left[
1-\alpha\right] ^{-1}\log\mathrm{Tr}\{  \rho_{A}^{\alpha}\} ,
\end{equation}
where the parameter $\alpha\in( 0,1) \cup (1,\infty)  $. (It is traditionally defined for $\alpha\in\left\{  0,1,\infty\right\}  $ in the limit as $\alpha$ approaches $0$, $1$, and $\infty$, respectively). The $\alpha$-R{\'e}nyi entropy is non-negative, additive on tensor-product states, and converges to the von Neumann entropy $H(A)_{\rho}\equiv-\mathrm{Tr}\{\rho_{A}\log\rho_{A}\}$ in the limit as the R{\'e}nyi parameter $\alpha$ tends to one. The $\alpha$-R{\'e}nyi entanglement entropies characterize the entanglement spectra of condensed matter systems, akin to moments of a probability distribution~\cite{PhysRevLett.101.010504, FHHW09}.
In Gaussian quantum information theory, the
$\alpha=2$ R{\'e}nyi entropy is useful in studying Gaussian entanglement and
other more general quantum correlations~\cite{AGS12}. In quantum
thermodynamics, $\alpha$-R{\'e}nyi entropies represent the derivative of the
free energy with respect to temperature~\cite{Baez_11} and are relevant for the
work value of information~\cite{Dahlsten_11}. In
quantum information theory, the von Neumann entropy captures
the amount of quantum information contained in an ensemble of a
large number of independent and identically distributed (i.i.d.)~copies
of the state \cite{PhysRevA.51.2738}, while the $\alpha$-R{\'e}nyi
entropies (for values of $\alpha$ other than one) are relevant in
scenarios beyond the i.i.d.~setting~\cite{T12}. For example,
the $\alpha$-R{\'e}nyi entropies are useful for characterizing
information processing tasks in regimes of a single or finite number
of resource utilizations~\cite{Holevo00, Hayashi07, MH11,MO13} and
for establishing strong converse
theorems~\cite{Arimoto73, ON99, KW09, PV10, SW12, WWY13,GW13,TWW14}. Given the rich variety of applications of the R{\'e}nyi entropies,
there has been a substantial effort towards
obtaining R{\'e}nyi generalizations of other information measures, such as the
conditional quantum entropy (CQE), or the quantum mutual information (QMI)~\cite{ON99,KW09,MH11,SW12,MDSFT13,WWY13,GW13,TBH13}. Recently, the authors of the present paper have contributed to this effort by proposing R\'enyi generalizations of the conditional quantum mutual information (CQMI)~\cite{BSW14}.

In this paper, we consider more generally a whole class of quantum information measures---we prescribe R{\'e}nyi generalizations for any quantum information measure which is equal to a linear
combination of von Neumann entropies with coefficients chosen from the set $\left\{
-1,0,1\right\} $. This criterion is met by many useful measures including the CQE, the QMI, and the CQMI, which are defined respectively as
\begin{align}
H(  A|B)  _{\rho}  &  \equiv H(  AB)
_{\rho}-H(  B)  _{\rho},\label{condent_1}\\
I(  A;B)  _{\rho}  &  \equiv H(  A)  _{\rho}+H(B)  _{\rho}-H(  AB)  _{\rho},\label{muinfo_1}\\
I( A;B|C) _{\rho}  & \equiv H( AC) _{\rho}+H( BC)
_{\rho}-H( C) _{\rho}\nonumber\\
&\ \ \ \ \ \ \ -H( ABC) _{\rho},\label{condmuinfo_1}
\end{align}
where $\rho$ is taken to be a bipartite state
$\rho_{AB}$ in (\ref{condent_1}) and (\ref{muinfo_1}), and a
tripartite state $\rho_{ABC}$ in (\ref{condmuinfo_1}). As examples, we describe R{\'e}nyi CQMIs, some quantum multipartite information measures and topological entanglement entropy (TEE) obtained based on the prescription. In particular, we focus on the R\'enyi CQMIs; we discuss the desired properties of the von Neumann CQMI that they retain, and some applications. We conjecture that the proposed R{\'e}nyi CQMIs are monotone increasing in the R{\'e}nyi parameter, with established proofs for some special cases. The solution of this conjecture would imply a characterization of states with CQMI nearly equal to zero, which could be helpful for solving some open problems in quantum information theory~\cite{Winterconj,Touchette14} and condensed matter physics~\cite{K13conj, K13thesis}.

The paper is organized as follows. In Section~\ref{bgrnd}, we set the stage by describing the earlier approaches to R\'enyi generalization of quantum information measures and their shortcomings. In Section~\ref{sec:genlincomb}, we present the new prescription for R\'enyi generalization of quantum information measures which are equal to a linear combination of von Neumann entropies with coefficients chosen from the set $\left\{-1,0,1\right\} $. In Section~\ref{examapp}, we describe examples, namely, R\'enyi CQMI, some R\'enyi multipartite information measures and R\'enyi TEE. Finally, we state our conclusions in Section~\ref{concl}.

\section{Background}
\label{bgrnd}

Suppose that we would like to establish a R\'{e}nyi generalization of the following
linear combination of entropies:
\begin{equation}
\sum_{S\subseteq\left\{  A_{1},\ldots,A_{l}\right\}  }a_{S}H(  S)
_{\rho},\label{eq:linear-combo-vN}%
\end{equation}
where $\rho_{A_{1}\ldots A_{l}}$ is a density operator on $l$ systems,
the coefficients $a_{S}\in\left\{  -1,0,1\right\}  $, and the sum runs over all subsets of the
systems $A_{1},\ldots,A_{l}$. A first approach one might consider is simply to replace the linear combination of von Neumann entropies
with the corresponding linear combination of $\alpha$-R{\'e}nyi
entropies:
\begin{equation}
\sum_{S\subseteq\left\{  A_{1},\ldots,A_{l}\right\}  }a_{S}H_{\alpha}\left(
S\right)  _{\rho}. \label{eq:naive-renyi}%
\end{equation}
However, the work of \cite{LMW13} establishes that there are no universal
constraints on such a quantity. For example, consider the following quantity obtained by replacing the von Neumann entropies in (\ref{condmuinfo_1}) with $\alpha$-R{\'e}nyi entropies, namely
\begin{align}
I_{\alpha}^{\prime}(A;B|C)_{\rho}&\equiv H_{\alpha}(AC)_{\rho}+H_{\alpha
}(BC)_{\rho}-H_{\alpha}(C)_{\rho}\nonumber\\
&\ \ \ \ \ \ \ -H_{\alpha}(ABC)_{\rho}.
\label{eq:simple-renyi-generalization}%
\end{align}
For $\alpha \in (0,1)\cup(1,\infty)$, this quantity does not generally satisfy non-negativity~\cite{LMW13}, while the von Neumann CQMI is known to be non-negative, the latter being a result of the strong subadditivity inequality~\cite{LR73}. Since strong subadditivity is consistently useful in applications and often
regarded as a \textquotedblleft law of quantum information
theory,\textquotedblright\ the work in \cite{LMW13}\ suggests that the
R\'{e}nyi generalization in (\ref{eq:naive-renyi}) is perhaps not the
appropriate one to be using in applications~\footnote{There are exceptions though, if we restrict ourselves to special types of quantum states. For example, for Gaussian states, the $\alpha=2$ R{\'e}nyi entropy satisfies strong subadditivity, as was shown in~\cite{AGS12}}.

On the other hand, 
one can write a quantum information measure in terms of the relative entropy, and subsequently replace the relative entropy with
a R{\'e}nyi relative entropy \cite{P86}, \cite{MDSFT13}, \cite{WWY13} in order to obtain a R{\'e}nyi generalization of the measure. For density operators $\rho$ and $\sigma$, the relative entropy is defined as 
\begin{equation}
D(\rho\Vert\sigma)\equiv\text{Tr}%
\left\{  \rho\log\rho\right\}  -\text{Tr}\left\{  \rho\log\sigma\right\}
\label{eq:vn-rel-ent}%
\end{equation}
if $\text{supp}\left(  \rho\right)  \subseteq\text{supp}\left(
\sigma\right)$, and it is equal to $+\infty$ otherwise.
The R{\'e}nyi relative entropy between density operators $\rho$ and $\sigma$ is defined
for $\alpha \in [0,1) \cup (1,\infty)$ 
as~\cite{P86}
\begin{equation}
D_{\alpha}(\rho\Vert\sigma)\equiv
\frac{1}{\alpha-1}\log\text{Tr}\left\{  \rho^{\alpha}\sigma^{1-\alpha}\right\}%
\label{eq:Renyi-rel-ent}%
\end{equation}
if $\text{supp}\left(  \rho\right)  \subseteq\text{supp}\left(  \sigma\right)$ or 
$(\alpha\in\lbrack0,1)\text{ and }\rho\not \perp \sigma)$. It is equal to 
$+\infty$ otherwise 
(these support conditions were established in \cite{TCR09}).
The von Neumann entropy of a state $\rho_{A}$ on system $A$ can itself be
written in terms of the relative entropy as $-D\left(  \rho_{A}\|I_{A}\right) $, where $I_{A}
$ is the identity operator. The CQE and the QMI
can also be written in terms of the relative entropy as
\begin{align}
H(  A|B)  _{\rho}  & 
= -\min_{\sigma_B}D\left(  \rho_{AB}\|I_{A}\otimes\sigma_{B}\right)  ,\label{renyiecem2}\\
I(  A;B)  _{\rho}  &  =
\min_{\sigma_B}D\left( \rho_{AB}\|\rho_{A}\otimes\sigma_{B}\right) ,\label{renyiecemu3}
\end{align}
respectively, where $\rho_{A}$ is the reduced density operator $\mathrm{Tr}_{B}\{\rho_{AB}\}$ and $\sigma_{B}$ is any density operator on the Hilbert space $\mathcal{H}_{B}$ of system $B$. (The unique optimum $\sigma_B$ in the above expressions turns out to be the reduced density operator $\rho_B$.)
Therefore, one can obtain R{\'e}nyi generalizations of the above quantities by using the R{\'e}nyi relative entropy in
place of the relative entropy. R\'{e}nyi generalizations of quantum
information measures obtained via the above procedure converge to the corresponding von Neumann entropy based quantities in the limit as $\alpha$ tends to one. They also retain most of the desired properties of the original quantities. For example, a R\'{e}nyi QMI obtained from~(\ref{renyiecemu3}), just like the original von Neumann entropy based quantity, is non-negative and non-increasing under the action of local completely positive and trace preserving (CPTP) maps for $\alpha\in\left[0,1\right)\cup\left(1, 2\right]$. This is because the R\'{e}nyi relative entropy for $\alpha\in\left[0,1\right)\cup\left(1, 2\right]$, just like the relative entropy, is non-negative and non-increasing under the action of any CPTP map, in the sense that
\begin{equation}
D_{\alpha}(  \rho\|\sigma)  \geq D_{\alpha}(  \mathcal{N}%
(  \rho)  \|\mathcal{N}(  \sigma)  )
\end{equation}
for a quantum map $\mathcal{N}$ \cite{P86}.

One could also use the sandwiched R\'{e}nyi
relative entropy \cite{MDSFT13}, \cite{WWY13}---a new variant of the R\'{e}nyi
relative entropy---instead of the R\'{e}nyi relative entropy. The sandwiched R\'{e}nyi relative entropy has found a number of applications in quantum information theory recently in the context of strong converse theorems~\cite{WWY13}, \cite{MO13}, \cite{GW13}, \cite{TWW14}. It is defined
for $\alpha \in (0,1) \cup (1,\infty)$
as
\begin{equation}
\widetilde{D}_{\alpha}\left(  \rho\Vert\sigma\right)\equiv \frac{1}{\alpha-1}\log\left[ \text{Tr}\left\{  \left(  \sigma^{\left(  1-\alpha\right)  /2\alpha}%
\rho\sigma^{\left(  1-\alpha\right)  /2\alpha}\right)  ^{\alpha}\right\}
\right]
\label{eq:def-sandwiched}%
\end{equation}
if $\text{supp}\left(  \rho\right)  \subseteq\text{supp}\left(  \sigma\right)$ or $\text{(}\alpha\in(0,1)\text{ and }\rho\not \perp \sigma\text{)}$. It is equal to $+\infty$ otherwise.
The sandwiched R\'{e}nyi relative entropy is non-negative and non-increasing under the action of any CPTP map for $\alpha\in\left[1/2,1\right)\cup\left(1, \infty\right) $ \cite{FL13}. Sandwiched R\'{e}nyi generalizations of quantum information measures as discussed above thus also satisfy the above properties.

In order to write an information quantity in terms of a relative entropy, the key task is to identify the second argument for the relative entropy. This task, however, can be nontrivial in some cases. For example, it is not obvious as to what the second argument should be for the CQMI. Taking a cue from
the QMI of~(\ref{renyiecemu3}), in which the second argument (when suitably normalized) has vanishing
QMI, one may try to write the CQMI as an
optimized relative entropy with respect to the set of quantum Markov states~\cite{HJPW04}, which are defined as those tripartite states which
have zero CQMI. However, it has been shown that
such a quantity is not equal to the CQMI, and can in general be arbitrarily large compared to the latter~\cite{ILW08}. Therefore, the relative entropy distance to the set of quantum Markov states does not lead to a good definition for R\'{e}nyi CQMI.

\section{Prescription for R\'enyi generalization}
\label{sec:genlincomb}

Having discussed the general approach towards obtaining R{\'e}nyi generalizations of
quantum information measures of the form given in (\ref{eq:linear-combo-vN}), and the hurdles faced, we now give our prescription for a R\'enyi generalization. It is also based on the relative entropy and its variants.

In the case that 
$a_{A_{1}\ldots A_{l}}$ is nonzero, without loss of generality,
we can set $a_{A_{1}\ldots A_{l}}=-1$ (otherwise,
factor out $-1$ to make this the case). Then, we can rewrite the quantity in (\ref{eq:linear-combo-vN})
in terms of the relative entropy as follows:
\begin{equation}\label{relentlin}
D\left(  \rho_{A_{1}\ldots A_{l}}\middle \|\exp\left\{
\sum_{S\subseteq A^{\prime}}a_{S}\log\rho_{S}\right\}  \right) ,
\end{equation}
where $A^{\prime}=\left\{  A_{1},\ldots,A_{l}\right\}  \backslash
A_{1}\cdots A_{l}$.
On the other hand, if $a_{A_{1}\ldots A_{l}}=0$, i.e., if all the marginal entropies in the sum are on a number of systems that is strictly smaller than the number of systems over which the state $\rho$ is defined (as is the case with $H(AB) + H(BC) +
H(AC)$, for example), we can take a purification of the original state
and call this purification the state
$\rho_{A_{1}\ldots A_{l}}$. This state is now a pure state on a number of systems
strictly larger than the number of systems involved in all the marginal
entropies. We then add the entropy $H(A_{1}\ldots A_{l})_{\rho}= 0$ to the sum
of entropies and apply the above recipe (so we resolve the issue with this
example by purifying to a system $R$, setting the sum formula to be $H(ABCR) +
H(AB) + H(BC) + H(AC)$, and proceeding with the above recipe). 

We then appeal to a
multipartite generalization of the Lie-Trotter product formula~\cite{S85}
to rewrite the second argument in~(\ref{relentlin}) as
\begin{equation}
\lim_{\alpha\rightarrow1}\left[ \underset{S\subseteq A^{\prime}}{\bigcirc}\Theta_{\rho
_{S}^{a_{S}\left(  1-\alpha\right)  /2}}\left(  I_{A_{1}\cdots A_{l}}\right)
\right]  ^{1/\left(  1-\alpha\right)  },\label{limLT}%
\end{equation}
where the map 
\begin{equation}
\Theta_{\rho_{S}^{a_{S}\left(  1-\alpha\right)  /2}}\left(
X\right)  \equiv\rho_{S}^{a_{S}\left(  1-\alpha\right)  /2}X\rho_{S}%
^{a_{S}\left(  1-\alpha\right)  /2}
\end{equation} 
and the composition $\bigcirc$ of maps
$\Theta_{\rho_{S}^{a_{S}\left(  1-\alpha\right)  /2}}$ for all subsets $S$ can
proceed in any order, and $I_{A_{1}\cdots
A_{l}}$ is the identity operator on the support of the state $\rho_{A_{1}\ldots A_{l}}$. Finally, we obtain a R\'{e}nyi generalization
of the linear combination in (\ref{eq:linear-combo-vN}) as%
\begin{equation}
D_{\alpha}\left(  \rho_{A_{1}\ldots A_{l}}\middle \|\left[   \underset{S\subseteq A^{\prime}}{\bigcirc}\Theta_{\rho_{S}^{a_{S}\left(  1-\alpha\right)  /2}}\left(  I_{A_{1}\cdots
A_{l}}\right)  \right]  ^{1/\left(  1-\alpha\right)  }\right)
,\label{eq:Renyi-gen-arbitrary}%
\end{equation}
where $D_{\alpha}$ is the R\'enyi relative entropy, and we have used (\ref{limLT}) in (\ref{relentlin}) and promoted the parameter $\alpha$ in~(\ref{limLT}) to take the role of the R\'{e}nyi parameter.

A similar R\'{e}nyi generalization can also be obtained using the sandwiched R\'enyi relative entropy as
\begin{equation}
\widetilde{D}_{\alpha}\left(  \rho_{A_{1}\ldots A_{l}}\middle \|\left[   \underset{S\subseteq A^{\prime}}{\bigcirc}\Theta_{\rho_{S}^{a_{S}\left(  1-\alpha\right)  /(2\alpha)}}\left(  I_{A_{1}\cdots
A_{l}}\right)  \right]  ^{\alpha/\left(  1-\alpha\right)  }\right).
\label{eq:Renyisanwi}%
\end{equation}
Note that there further exist a number of different possible variants of the above R\'{e}nyi generalizations since different choice of orderings of the maps $\Theta_{\rho_{S}^{a_{S}\left(  1-\alpha\right)  /2}}$ are possible. Moreover, we could also consider arbitrary density operators on the appropriate subsystems for the maps $\Theta$ instead of the reduced density operators (marginals) of $\rho_{A_{1}\ldots A_{l}}$, and then optimize over these operators (under the assumption that the support of $\rho_{A_{1}\ldots A_{l}}$ is contained in the intersection of the supports of these operators).

For any quantum information measure, it is possible to prove that these different R\'enyi generalizations converge to the original von Neumann entropy based quantity in \eqref{eq:linear-combo-vN} in the limit as $\alpha\rightarrow1$. Also, consider that we can write
the linear combination in (\ref{eq:linear-combo-vN}) as%
\begin{align}
&\sum_{S\subseteq\left\{  A_{1},\ldots,A_{l}\right\}  }a_{S}H\left(  S\right)
_{\rho}\nonumber\\
&=-\text{Tr}\left\{  \rho_{A_{1}\ldots A_{l}}\left[  \sum_{S\subseteq
\left\{  A_{1},\ldots,A_{l}\right\}  }a_{S}\log\rho_{S}\right]  \right\}  .
\end{align}
The \textit{information second moment} corresponding to this combination of
entropies is then%
\begin{align}
&V\left(  \rho_{A_{1}\ldots A_{l}},\left\{  a_{S}\right\}  \right)\nonumber\\
&\equiv\text{Tr}\left\{  \rho_{A_{1}\ldots A_{l}}\left[  \sum_{S\subseteq
\left\{  A_{1},\ldots,A_{l}\right\}  }a_{S}\log\rho_{S}\right]  ^{2}\right\}.
\end{align}
It can be shown that the R\'{e}nyi generalization in (\ref{eq:Renyi-gen-arbitrary}) has
the following Taylor expansion about $\gamma=0$, where $\gamma=\alpha-1$:%
\begin{align}
&\frac{1}{\gamma}\log\bigg[  \text{Tr}\left\{  \rho_{A_{1}\ldots A_{l}%
}\right\}  +\gamma\sum_{S\subseteq\left\{  A_{1},\ldots,A_{l}\right\}  }%
a_{S}H\left(  S\right)  _{\rho}\nonumber\\
&+\frac{\gamma^{2}}{2}V\left(  \rho_{A_{1}\ldots
A_{l}},\left\{  a_{S}\right\}  \right)  +O\left(  \gamma^{3}\right)\bigg],
\end{align}
thus recovering the information second moment as the second order term in the
Taylor expansion. (See~\cite[Appendix E.1]{BSW14}, for example, which shows the Taylor expansion in a neighborhood of $\gamma=0$ for the R\'enyi CQMI.) Note that the R\'{e}nyi generalization in
(\ref{eq:naive-renyi}) does not recover the information second moment in a
Taylor expansion. Furthermore, we leave it as an open question to determine whether the following statement is generally true: \textit{if a von Neumann entropy based measure is non-negative and non-increasing under the action of local CPTP maps, then its R\'{e}nyi generalizations of the above type are also non-negative and non-increasing under local CPTP maps}.

\section{Examples of R\'enyi quantum information measures}
\label{examapp}

\subsection{R\'enyi conditional quantum mutual information}

The CQMI quantifies how much correlation exists in a
tripartite state between two parties from the perspective of the
third. A compelling operational interpretation of the quantity
has been given in terms of the quantum state redistribution
protocol \cite{DY08,YD09}:\ given a four-party pure state
$\psi_{ADBC}$, with a sender possessing systems $D$ and $B$ and a receiver
possessing system $C$, the optimal rate of quantum communication necessary to
transfer the system $B$ to the receiver is given by $\tfrac{1}{2}I(
A;B|C)  _{\psi}$. As mentioned before, due to strong subadditivity~\cite{LR73}, the
CQMI is non-negative. Additionally, also due to
strong subadditivity, the CQMI can never increase under local CPTP maps performed on the systems $A$ or $B$ \cite{CW04}, so that $I(
A;B|C)  _{\rho}$ is a sensible measure of the correlations present
between systems $A$ and $B$, from the perspective of $C$. That is, the
following inequality holds%
\begin{equation}
I(  A;B|C)  _{\rho}\geq I(  A';B'|C)  _{\omega},
\end{equation}
where $\omega_{A'B'C}\equiv\left(  \mathcal{N}_{A\rightarrow A'}\otimes\mathcal{M}_{B\rightarrow B'}\right)
\left(  \rho_{ABC}\right)  $ with $\mathcal{N}_{A\rightarrow A'}$ and $\mathcal{M}_{B\rightarrow B'}$
arbitrary local CPTP maps performed on the systems $A$ and $B$,
respectively. One other property of the CQMI is that it obeys a
duality relation~\cite{DY08,YD09}. That is, for a four-party pure state
$\psi_{ABCD}$, the following equality holds%
\begin{equation}
I(  A;B|C)  _{\psi}=I(  A;B|D)  _{\psi}.\label{Duality}%
\end{equation} 

The CQMI finds numerous applications. In quantum many body physics, the quantum Markov states (tripartite
states which have zero CQMI) can be readily used to study gapped systems in two spatial dimensions~\cite{K13thesis}. The CQMI can also be used to study the mediation of quantum correlations in condensed matter systems~\cite{B12}. In entanglement theory, the CQMI underlies the squashed entanglement~\cite{CW04, KW04, AF04, BCY11}, which is the only measure of entanglement known to satisfy all axioms for an entanglement measure. The squashed entanglement of a state is an upper bound on the amount of entanglement that can be extracted from the state by an entanglement distillation protocol~\cite{CW04}.
Furthermore, the squashed entanglement of a channel is known to be
an upper bound on the quantum communication capacity of any channel assisted
by unlimited forward and backward classical communication~\cite{TGW13}. The CQMI also underlies 
quantum discord~\cite{Pia12}. Quantum discord
is a measure of quantum correlations different from entanglement \cite{zurek}
and has been studied quite extensively~\cite{KBCPV12}.

\begin{table*}
\begin{tabular}{|c|c|c|c|c|}
\hline 
Formula & CQMI in (\ref{condmuinfo_1}) & R\'enyi CQMI in (\ref{eq:simple-renyi-generalization}) & R\'enyi CQMI in (\ref{sibsonRCQMI}) & R\'enyi CQMI in (\ref{renyicmiopt}) \tabularnewline
\hline 
\hline 
Non-negative & \Checkmark{} & \XSolid{} & \Checkmark{} & \Checkmark{}\tabularnewline
\hline 
Monotone under local op.'s on A & \Checkmark{} & \XSolid{} & ? & ?\tabularnewline
\hline 
Monotone under local op.'s on B & \Checkmark{} & \XSolid{} & \Checkmark{} & \Checkmark{}\tabularnewline
\hline 
Duality & \Checkmark{} & \Checkmark{} & \Checkmark{} & \Checkmark{}\tabularnewline
\hline 
Converges to (\ref{condmuinfo_1})
as $\alpha \to 1$ & N/A & \Checkmark{} & \Checkmark & ?\tabularnewline
\hline 
Monotone in $\alpha$ & N/A & \XSolid{} & ? & ?\tabularnewline
\hline 
\end{tabular}
\caption{R\'enyi generalizations of the conditional quantum mutual information (CQMI). The R\'{e}nyi generalizations prescribed in this work are applicable to the CQMI. The leftmost column of the table lists some
 desired properties of a R\'enyi CQMI. These properties are satisfied by the original von Neumann CQMI $I(A;B|C)_\rho$ in (\ref{condmuinfo_1}) as shown in Column~2. The R\'enyi CQMI in (\ref{eq:simple-renyi-generalization}) obtained by simply replacing the linear sum of von Neumann entropies with the corresponding linear sum of R\'enyi entropies, in Column 3, is compared with the R\'enyi generalizations obtained through the formula prescribed in this work, in Columns 4 and 5. The question marks indicate open questions, with numerical evidence supporting a positive answer. The quantity in Column~3 does not retain many of the desired properties. On the contrary, the quantities in Columns~4 and 5 retain some of these desired properties. The table suggests that the latter are more useful R\'enyi generalizations of the CQMI.}
\label{sumtab}
\end{table*}

Using the prescribed formula of (\ref{eq:Renyi-gen-arbitrary}), a R\'enyi CQMI can be defined as
\begin{align}
I_\alpha(A;B|C)  _{\rho}&\equiv \underset{\sigma_{BC}}{\min} \ D_{\alpha
}\left(  \rho_{ABC}\Vert Y(\alpha)\right)\nonumber\\
& =\underset{\sigma_{BC}}{\min}\frac{1}{\alpha-1}\log{\rm Tr}\left\{\rho_{ABC}^\alpha Y(\alpha)^{1-\alpha}\right\},
\end{align}
where $D_\alpha$ is the R\'enyi relative entropy of order $\alpha$ given in (\ref{eq:Renyi-rel-ent}), and
\begin{equation}
Y(\alpha)\equiv\left[  \rho_{AC}^{\left(  1-\alpha\right)  /2}\rho
_{C}^{\left(  \alpha-1\right)  /2}\sigma_{BC}^{1-\alpha}\rho_{C}^{\left(
\alpha-1\right)  /2}\rho_{AC}^{\left(  1-\alpha\right)  /2}\right]
^{1/\left(  1-\alpha\right)  }.
\end{equation}
The choice of minimization over the $BC$ subsystem enables the duality property of~(\ref{Duality}) to hold for the resulting quantity~\cite[Theorem 32]{BSW14}. Using a Sibson-like identity~\cite{S69}, the unique optimum state $\sigma_{BC}$ can be determined explicitly, and the above R\'{e}nyi CQMI can be rewritten as
\begin{equation}
I_\alpha(A;B|C)  _{\rho}=\frac{\alpha}{\alpha-1}\log
\operatorname{Tr}\left\{  Z(\alpha)\right\},
\label{sibsonRCQMI}
\end{equation}
where $Z(\alpha)$ is
\begin{equation}
\left(  \rho_{C}^{\left(  \alpha-1\right)
/2}\operatorname{Tr}_{A}\left\{  \rho_{AC}^{\left(  1-\alpha\right)  /2}%
\rho_{ABC}^{\alpha}\rho_{AC}^{\left(  1-\alpha\right)  /2}\right\}  \rho
_{C}^{\left(  \alpha-1\right)  /2}\right)  ^{1/\alpha}
\end{equation}
\cite[Proposition 8]{BSW14}. The above quantity converges to the von Neumann CQMI in
the limit as $\alpha\rightarrow1$ for any tripartite state $\rho_{ABC}$ \cite[Theorem 11]{BSW14}. It is non-negative and non-increasing under the action of local CPTP maps on the $B$ system in the
range of $\alpha\in\left[0,1\right)\cup\left(1, 2\right]$~\cite[Corollaries 16 and 15]{BSW14};
it is an open question to determine whether \textit{the
quantity is also non-increasing under local CPTP maps on the $A$ system}
(numerical work has supported a positive answer). We can also define a sandwiched R\'{e}nyi
conditional mutual information $\widetilde{I}_{\alpha}(A;B|C)_{\rho}$ using
the sandwiched R\'{e}nyi relative entropy $\widetilde{D}_\alpha$. A particular sandwiched R\'{e}nyi CQMI that satisfies the desired properties of non-negativity, monotonicity under local CPTP maps on
the $B$ system (in the range of $\alpha\in\left[1/2,1\right)\cup\left(1, \infty\right) $), and duality \cite[Section~6]{BSW14} is given by
\begin{align}
\label{renyicmiopt}
\tilde{I}_\alpha(A;B|C)  _{\rho}&\equiv \underset{\sigma_{BC}}{\inf} \underset{\omega_{C}}{\sup}\ \tilde{D}_{\alpha
}\left(  \rho_{ABC}\Vert Y'(\alpha)\right),
\end{align}
where $\tilde{D}_{\alpha}$ is the sandwiched R\'enyi relative entropy of order $\alpha$ given in (\ref{eq:def-sandwiched}), and 
\begin{multline}
Y'(\alpha) \equiv\bigg[  \rho_{AC}^{\left(  1-\alpha\right)  /(2\alpha)}\omega_{C}^{\left(
\alpha-1\right)  /(2\alpha)}\sigma_{BC}^{(1-\alpha)/\alpha}\\
\times\omega_{C}^{\left(  \alpha-1\right)
/(2\alpha)}\rho_{AC}^{\left(  1-\alpha\right)  /(2\alpha)}\bigg]  ^{\alpha/\left(  1-\alpha
\right)  }.
\end{multline}

Table~\ref{sumtab} summarizes the various properties of the above-described R\'enyi and sandwiched R\'enyi CQMIs in comparison with the R\'enyi CQMI given in (\ref{eq:simple-renyi-generalization}). This includes the property of monotonicity of these quantities in the R\'enyi parameter, which we discuss later in Section~\ref{monalphaconj}. The comparison elucidates the effectiveness of the prescribed formula. The R{\'e}nyi CQMIs also converge to the quantity in (\ref{condmuinfo_1}) in the limit as the R{\'e}nyi parameter tends to one.

\subsubsection{R\'{e}nyi squashed entanglement}

The squashed entanglement of a bipartite state $\rho_{AB}$\ is defined as
\cite{CW04}%
\begin{align}
&E_{\text{sq}}\left(  A;B\right)  _{\rho}\nonumber\\
&\equiv\frac{1}{2}\inf_{\rho_{ABE}%
}\left\{  I\left(  A;B|E\right)  _{\rho}:\rho_{AB}=\text{Tr}_{E}\left\{
\rho_{ABE}\right\}  \right\}  .
\end{align}
Thus, a straightforward R\'{e}nyi generalization is as follows:%
\begin{align}
&E_{\text{sq}}^{\alpha}\left(  A;B\right)  _{\rho}\nonumber\\
&\equiv\frac{1}{2}\inf
_{\rho_{ABE}}\left\{  I_{\alpha}\left(  A;B|E\right)  _{\rho}:\rho
_{AB}=\text{Tr}_{E}\left\{  \rho_{ABE}\right\}  \right\}  .
\end{align}
A R\'enyi squashed entanglement could potentially be used to strengthen the results on distillable entanglement of a state and the quantum communication capacity of a channel assisted by unlimited two-way classical communication by establishing the squashed entanglement as a strong converse rate for these respective tasks. It remains a topic of future research to investigate these applications of the R\'enyi squashed entanglement in full detail. Also, it is important to establish that the R\'{e}nyi conditional mutual informations satisfy monotonicity under local quantum operations on both $A$ and $B$ in order for this quantity to be a sensible correlation measure. Assuming the truth of the above statement (with the support of numerical evidence), in a followup work, we have explored the various properties of the R\'enyi squashed entanglement that would qualify it as an entanglement monotone~\cite[Section 4]{SBW14}. 

\subsubsection{R\'{e}nyi quantum discord}

We obtain a R\'{e}nyi generalization of the quantum discord \cite{zurek},
since we can write the discord of a bipartite state $\rho_{AB}$ as \cite{Pia12}%
\begin{align}
& \!\!\! I\left(  A;B\right)  -\sup_{\left\{  \Lambda\right\}  }I\left(  X;B\right) \nonumber \\
&  =\inf_{\left\{  \Lambda\right\}  }I\left(  A;B\right)  -I\left(  X;B\right)
\\
&  =\inf_{\left\{  \Lambda\right\}  }I\left(  EX;B\right)  -I\left(
X;B\right) \\
&  =\inf_{\left\{  \Lambda\right\}  }I\left(  E;B|X\right)  ,
\end{align}
where the optimization is over all POVMs acting on the system $A$, with
classical output $X$. We can then find an isometric extension of any such
measurement, where we label the environment system as $E$. So we just define a
R\'{e}nyi quantum discord in the following way:%
\begin{equation}
\inf_{\left\{  \Lambda\right\}  }I_{\alpha}\left(  E;B|X\right)  ,
\end{equation}
where $I_{\alpha}$ could be any R\'{e}nyi generalization of the conditional
mutual information. Such a R\'enyi discord might potentially bear new insights over the von Neumann discord~\cite[Section 5]{SBW14}.

\subsubsection{Potential operational characterization of states with small CQMI}
\label{monalphaconj}

The R\'{e}nyi CQMIs could also be useful in connection with some important open problems in quantum information theory and condensed matter physics. We conjecture that {\it our R\'{e}nyi  generalizations of the CQMI are monotonically increasing
functions of the R\'{e}nyi parameter}. That is, for $0\leq\alpha\leq\beta$, we conjecture that
\begin{align}
I_{\alpha}(A;B|C)_{\rho}\leq I_{\beta}(A;B|C)_{\rho}, \label{eq:conj-1}
\end{align}
as well as the analogous statement for the sandwiched R\'{e}nyi CQMI. These conjectures are true in a number of special cases. For example, we it can be shown that these conjectures hold when the R\'{e}nyi parameter $\alpha$ is in a neighborhood of one, and that \eqref{eq:conj-1}
is true in the case when $\alpha+\beta=2$ (or when $1/\alpha+1/\beta=2$ for the sandwiched R\'{e}nyi CQMI) \cite[Section 8]{BSW14}. If proven to be correct generally, the above conjecture would establish the truth of an open conjecture from \cite{K13conj} (up to a constant):
\begin{align}
I(A;B|C)_{\rho} & \geq \widetilde{I}_{1/2}(A;B|C)_{\rho}\\
& =-\log F(\rho_{ABC},\mathcal{R}(\rho_{BC})),
\label{fawren}
\end{align}
where $\mathcal{R}(\cdot)\equiv\rho_{AC}^{1/2}\rho_{C}^{-1/2}%
(\cdot)\rho_{C}^{-1/2}\rho_{AC}^{1/2}$ denotes Petz's recovery map for the partial trace over $A$~\cite{Petz93} and $F(P,Q)\equiv \Vert\sqrt{P}\sqrt{Q} \Vert_1^2$ is the quantum fidelity. This would give an operational characterization of quantum states with small CQMI (i.e., states that fulfill strong subadditivity with near equality). This characterization could be useful for
understanding topological order in condensed matter physics~\cite{K13conj,K13thesis}, for solving some open questions related to squashed entanglement~\cite{Winterconj}, as well as for deriving quantum communication complexity lower bounds~\cite{KNTZ01}, as discussed in~\cite{Touchette14}. After the completion of the present paper,
recent progress on establishing (\ref{fawren}) has appeared in \cite{FR14}.

\subsection{R\'{e}nyi multipartite information measures}

\label{sec:multipartite-CMI}The conditional multipartite information of a
state $\rho_{A_{1}\cdots A_{l}C}$ is defined as%
\begin{align}
&I\left(  A_{1};A_{2};\cdots;A_{l}|C\right)  _{\rho} \nonumber\\
& \equiv\left[\sum_{i=1}^{l}H\left(  A_{i}|C\right)  _{\rho}\right]  -H\left(  A_{1}%
A_{2}\cdots A_{l}|C\right)  _{\rho}\\
&=\left(  \sum_{i=1}^{l}\left[  H\left(  A_{i}C\right)  _{\rho}-H\left(
C\right)  _{\rho}\right]  \right)\nonumber\\
&-H\left(  A_{1}A_{2}\cdots A_{l}C\right)_{\rho}+H\left(  C\right)  _{\rho}%
\end{align}
and has appeared in various contexts (see \cite{YHHHOS09}\ and references
therein). This quantity can be written as a relative entropy as follows:%
\begin{align}
&I\left(  A_{1};A_{2};\cdots;A_{l}|C\right)  _{\rho}\nonumber\\
&=D\left(  \rho_{A_{1}\cdots
A_{l}C}\middle\Vert\exp\left\{  \sum_{i=1}^{l}\log\rho_{A_{i}C}-\sum
_{i=1}^{l-1}\log\rho_{C}\right\}  \right).
\end{align}
Thus, we obtain two natural R\'{e}nyi generalizations of this measure, defined
as%
\begin{align}
&I_{\alpha}\left(  A_{1};A_{2};\cdots;A_{l}|C\right)  _{\rho|\rho} \nonumber\\
&\equiv\frac{1}{\alpha-1}\log\text{Tr}\left\{  \rho_{A_{1}\cdots A_{l}%
C}^{\alpha}\sigma_{A_{1}\cdots A_{l}C}\left(  \alpha\right)  \right\}  \label{Renyi-C-mult-I},\\
&\widetilde{I}_{\alpha}\left(  A_{1};A_{2};\cdots;A_{l}|C\right)  _{\rho|\rho}\nonumber\\
&  \equiv\frac{1}{\alpha-1}\log\text{Tr}\left\{  \left(  \rho_{A_{1}\cdots
A_{l}C}^{1/2}\omega_{A_{1}\cdots A_{l}C}\left(  \alpha\right)  \rho
_{A_{1}\cdots A_{l}C}^{1/2}\right)  ^{\alpha}\right\}  \label{Sand-R-C-mult-I},
\end{align}
where%
\begin{align}
\sigma_{A_{1}\cdots A_{l}C}\left(  \alpha\right)&  \equiv\rho_{A_{l}%
C}^{\frac{1-\alpha}{2}}\rho_{C}^{\frac{\alpha-1}{2}}\cdots\rho_{C}%
^{\frac{\alpha-1}{2}}\rho_{A_{2}C}^{\frac{1-\alpha}{2}}\rho_{C}^{\frac
{\alpha-1}{2}}\rho_{A_{1}C}^{1-\alpha}\nonumber\\
&\times\rho_{C}^{\frac{\alpha-1}{2}}\rho
_{A_{2}C}^{\frac{1-\alpha}{2}}\rho_{C}^{\frac{\alpha-1}{2}}\cdots\rho
_{C}^{\frac{\alpha-1}{2}}\rho_{A_{l}C}^{\frac{1-\alpha}{2}},\\
\omega_{A_{1}\cdots A_{l}C}\left(  \alpha\right)&   \equiv\rho_{A_{l}%
C}^{\frac{1-\alpha}{2\alpha}}\rho_{C}^{\frac{\alpha-1}{2\alpha}}\cdots\rho
_{C}^{\frac{\alpha-1}{2\alpha}}\rho_{A_{2}C}^{\frac{1-\alpha}{2\alpha}}%
\rho_{C}^{\frac{\alpha-1}{2\alpha}}\rho_{A_{1}C}^{\frac{1-\alpha}{\alpha}}\nonumber\\%
&\times\rho_{C}^{\frac{\alpha-1}{2\alpha}}\rho_{A_{2}C}^{\frac{1-\alpha}{2\alpha}%
}\rho_{C}^{\frac{\alpha-1}{2\alpha}}\cdots\rho_{C}^{\frac{\alpha-1}{2\alpha}%
}\rho_{A_{l}C}^{\frac{1-\alpha}{2\alpha}}.
\end{align}
These quantities satisfy monotonicity under local quantum operations on the
system $A_{1}$ for $\alpha\in\lbrack0,1)\cup(1,2]$ and $\alpha\in
\lbrack1/2,1)\cup(1,\infty)$, respectively, and they reduce to the von Neumann
entropy in the limit as $\alpha\rightarrow1$. The proofs for these facts
follow along similar lines given in previous sections of this paper. It is an
open question to determine if the monotonicity holds with respect to local
operations on the individual systems $A_{1}$, \ldots, $A_{l}$.

For a state $\rho_{A_{1}A_{1}^{\prime}\cdots A_{l}A_{l}^{\prime}}$, another
multipartite information-based quantity that has appeared in
various contexts (see \cite{YHW08, PHH08, W14}\ and references therein) is as
follows:%
\begin{equation}
I\left(  A_{1}A_{1}^{\prime}:\cdots:A_{l}A_{l}^{\prime}\right)  -I\left(
A_{1}^{\prime}:\cdots:A_{l}^{\prime}\right)  .
\end{equation}
The measure gives an alternate approach for ``quantum conditioning,''
as discussed in \cite{YHW08}.
It can be written as the following relative entropy:%
\begin{multline}
D\bigg(\rho_{A_{1}A_{1}^{\prime}\cdots A_{l}A_{l}^{\prime}}\bigg\Vert\exp
\bigg\{\log\rho_{A_{1}^{\prime}\cdots A_{l}^{\prime}}+\\
\sum_{i=1}^{l}\log\rho_{A_{i}A_{i}^{\prime}}-\log\rho_{A_{i}^{\prime}%
}\bigg\}\bigg).
\end{multline}
Thus, we obtain two natural R\'{e}nyi generalizations of this quantity,
defined as%
\begin{align}
&  \frac{1}{\alpha-1}\log\text{Tr}\left\{  \rho_{A_{1}A_1'\cdots A_{l}A_l'}^{\alpha
}\sigma_{A_{1}\cdots A_{l}}\left(  \alpha\right)  \right\}  ,\\
&  \frac{1}{\alpha-1}\log\text{Tr}\left\{  \left(  \rho_{A_{1}A_1'\cdots A_{l}A_l'%
}^{1/2}\omega_{A_{1}\cdots A_{l}}\left(  \alpha\right)  \rho_{A_{1}A_1'\cdots
A_{l}A_l'}^{1/2}\right)  ^{\alpha}\right\}  ,
\end{align}
where%
\begin{multline}
\sigma_{A_{1}\cdots A_{l}C}\left(  \alpha\right)  \equiv\rho_{A_{1}^{\prime
}\cdots A_{l}^{\prime}}^{\frac{1-\alpha}{2}}\left(  \rho_{A_{1}^{\prime}%
}^{\frac{\alpha-1}{2}}\otimes\cdots\otimes\rho_{A_{l}^{\prime}}^{\frac
{\alpha-1}{2}}\right) \times\\ \left(  \rho_{A_{1}A_{1}^{\prime}}^{1-\alpha}%
\otimes\cdots\otimes\rho_{A_{l}A_{l}^{\prime}}^{1-\alpha}\right)
\left(  \rho_{A_{1}^{\prime}}^{\frac{\alpha-1}{2}}\otimes\cdots\otimes
\rho_{A_{l}^{\prime}}^{\frac{\alpha-1}{2}}\right)  \rho_{A_{1}^{\prime}\cdots
A_{l}^{\prime}}^{\frac{1-\alpha}{2}},
\end{multline}%
\begin{multline}
\omega_{A_{1}\cdots A_{l}C}\left(  \alpha\right)  \equiv\rho_{A_{1}^{\prime
}\cdots A_{l}^{\prime}}^{\frac{1-\alpha}{2\alpha}}\left(  \rho_{A_{1}^{\prime
}}^{\frac{\alpha-1}{2\alpha}}\otimes\cdots\otimes\rho_{A_{l}^{\prime}}%
^{\frac{\alpha-1}{2\alpha}}\right)  \times \\
\left(  \rho_{A_{1}A_{1}^{\prime}}%
^{\frac{1-\alpha}{\alpha}}\otimes\cdots\otimes\rho_{A_{l}A_{l}^{\prime}%
}^{\frac{1-\alpha}{\alpha}}\right)  
\left(  \rho_{A_{1}^{\prime}}^{\frac{\alpha-1}{2\alpha}}\otimes\cdots
\otimes\rho_{A_{l}^{\prime}}^{\frac{\alpha-1}{2\alpha}}\right)  \rho
_{A_{1}^{\prime}\cdots A_{l}^{\prime}}^{\frac{1-\alpha}{2\alpha}},
\end{multline}
These quantities satisfy monotonicity under local quantum operations on the
individual systems $A_{1}$, \ldots, $A_{l}$ for $\alpha\in\lbrack
0,1)\cup(1,2]$ and $\alpha\in\lbrack1/2,1)\cup(1,\infty)$, respectively, and
they reduce to the von Neumann entropy in the limit as $\alpha\rightarrow1$.
Proofs for these facts follow along similar lines as the proofs in
\cite[Sections 5 and 6]{BSW14}.

\subsection{R\'{e}nyi topological entanglement entropy}

\label{sec:TEE}The topological entanglement entropy of a tripartite quantum
state $\rho_{ABC}$\ is defined as the following linear combination of marginal
entropies \cite{KP06,PhysRevLett.96.110405}:%
\begin{align}
H_{\text{topo}}\left(  \rho_{ABC}\right) & \equiv H\left(  A\right)  _{\rho
}+H\left(  B\right)  _{\rho}+H\left(  C\right)  _{\rho}-H\left(  AB\right)_{\rho}\nonumber\\
&-H\left(  BC\right)  _{\rho}-H\left(  AC\right)  _{\rho}+H\left(
ABC\right)  _{\rho}.
\end{align}
This quantity is also known in the classical information theory literature as
the interaction information \cite{M54}. In condensed matter systems, the
topological entanglement entropy measures the long range quantum correlations
present in a many-body quantum state---it is helpful in classifying if and the
degree to which a system is topologically ordered. We can write
$H_{\text{topo}}\left(  \rho_{ABC}\right)  $ in terms of the relative entropy
as%
\begin{align}
-D\bigg(  \rho_{ABC}&\Vert\exp\big\{  \log\rho_{AB}+\log\rho_{BC}+\log
\rho_{AC}\nonumber\\
&-\log\rho_{A}-\log\rho_{B}-\log\rho_{C}\big\}  \bigg)  .
\end{align}
So this suggests one R\'{e}nyi generalization $H_{\text{topo}}^{\alpha}\left(
\rho_{ABC}\right)  $\ of the topological entanglement entropy to be%
\begin{align}
&-\frac{1}{\alpha-1}\log\text{Tr}\bigg\{  \rho_{ABC}^{\alpha}\rho_{AC}%
^{\frac{1-\alpha}{2}}\rho_{BC}^{\frac{1-\alpha}{2}}\rho_{AB}^{\frac{1-\alpha
}{2}}\rho_{C}^{\frac{\alpha-1}{2}}\rho_{B}^{\frac{\alpha-1}{2}}\rho
_{A}^{\alpha-1}\nonumber\\
&\times\rho_{B}^{\frac{\alpha-1}{2}}\rho_{C}^{\frac{\alpha-1}{2}}%
\rho_{AB}^{\frac{1-\alpha}{2}}\rho_{BC}^{\frac{1-\alpha}{2}}\rho_{AC}%
^{\frac{1-\alpha}{2}}\bigg\}  , \label{eq:renyi-topo-1}%
\end{align}
and a sandwiched R\'{e}nyi generalization $\widetilde{H}_{\text{topo}}%
^{\alpha}\left(  \rho_{ABC}\right)  $\ is as follows:%
\begin{align}
&-\frac{1}{\alpha-1}\log\text{Tr}\bigg\{  \big(  \rho_{ABC}^{1/2}\rho
_{AC}^{\frac{1-\alpha}{2\alpha}}\rho_{BC}^{\frac{1-\alpha}{2\alpha}}\rho
_{AB}^{\frac{1-\alpha}{2\alpha}}\rho_{C}^{\frac{\alpha-1}{2\alpha}}\rho
_{B}^{\frac{\alpha-1}{2\alpha}}\rho_{A}^{\frac{\alpha-1}{\alpha}}\nonumber\\
&\times\rho
_{B}^{\frac{\alpha-1}{2\alpha}}\rho_{C}^{\frac{\alpha-1}{2\alpha}}\rho
_{AB}^{\frac{1-\alpha}{2\alpha}}\rho_{BC}^{\frac{1-\alpha}{2\alpha}}\rho
_{AC}^{\frac{1-\alpha}{2\alpha}}\rho_{ABC}^{1/2}\big)  ^{\alpha}\bigg\}  .
\label{eq:renyi-topo-2}%
\end{align}
Of course, there are many generalizations depending on the ordering of the
operators and whether we use optimizations as discussed before.

In prior work, researchers had suggested that the entire entanglement spectrum
of a many-body state could lead to a finer classification of topological order
\cite{PhysRevLett.101.010504}. Following this proposal, \cite{FHHW09}
considered a R\'{e}nyi generalization of the topological entanglement entropy
along the lines in (\ref{eq:naive-renyi}). However, their conclusion was that
this particular R\'{e}nyi generalization does not lead to any further
universal information about topological order than that already provided by
the von Neumann topological entanglement entropy. A natural question to
consider going forward from here is whether the R\'{e}nyi topological
entanglement entropies as defined in (\ref{eq:renyi-topo-1}) and
(\ref{eq:renyi-topo-2}) could lead to extra desired universal information
about topological order in any setting (that is, whether it would depend on the R\'{e}nyi parameter
$\alpha$ in addition to the \textquotedblleft quantum
dimension\textquotedblright).

\section{Conclusion}
\label{concl}

We prescribed R\'enyi generalizations for quantum information measures
which consist of a linear combination of von Neumann entropies with
coefficients chosen from the set $\left\{-1,0,1\right\}$. The said
criterion is met by many useful quantum information measures such as
the conditional entropy, the mutual information, the conditional mutual
information, the conditional multipartite information, and the topological
entanglement entropy. While R\'enyi generalizations of the conditional entropy
and the mutual information were already known, it had been an open problem before
this work to obtain valid R\'enyi generalizations of the remaining
quantities listed above, in particular, of the conditional mutual information.

The R\'enyi generalizations of the prescribed type for the conditional
mutual information, $I_\alpha(A;B|C)$ satisfy many of the desired
properties of the original von Neumann quantity. They are non-negative,
monotone under local operations on either system $A$ or system $B$ (it
is left as an open question to prove the property for local operations
on the other system, but we have numerical evidence that supports a
positive answer), and obey a duality relation. Based on the R\'enyi
conditional mutual informations, we defined R\'enyi squashed entanglement
and R\'enyi discord and discussed some potential applications of these
quantities. We have explored these quantities in much more detail in our
followup work \cite{SBW14}. We also posed another open question about the proposed
R\'enyi conditional mutual informations being monotone increasing in
the R\'enyi parameter (we have proofs for this in some special cases
and numerical evidence supporting a positive answer in general), and
indicated an important implication of the conjecture in characterizing
quantum states with small conditional mutual information. Further,
we defined a R\'enyi conditional multipartite information and a R\'enyi
topological entanglement entropy. The R\'{e}nyi topological entanglement
entropies defined in (\ref{eq:renyi-topo-1}) and
(\ref{eq:renyi-topo-2}) present potentially interesting, new
possibilities in the study of topologically-ordered systems.

In view of the numerous potential uses of R\'{e}nyi generalizations of quantum information measures, our prescription for R\'{e}nyi generalization promises new avenues of research in both quantum information theory and other areas of physics.

\textbf{Note:} After the completion of this work, we learned of the recent breakthrough
result of \cite{FR14}, in which a variant of the inequality in (\ref{fawren}) has been
proven. This result has been further strengthened in~\cite{BHOS14,BLW14}, and its implications to robustness of quantum Markov chains and to other contexts have been further elucidated in~\cite{LW14,W14}.

\textbf{Acknowledgments.} MB\ is grateful for the hospitality of the Hearne
Institute for Theoretical Physics and the Department of Physics and Astronomy
at LSU\ for hosting him as a visitor during March 2014, when some of the
research in this paper was completed. KS acknowledges support from the Army
Research Office and NSF Grant No.~CCF-1350397. MMW\ acknowledges startup funds from the Department of
Physics and Astronomy at LSU, support from the NSF\ under Award
No.~CCF-1350397, and support from the DARPA Quiness Program through US Army
Research Office award W31P4Q-12-1-0019.

\bibliographystyle{apsrev4-1}
\bibliography{Ref_shortversion}

\end{document}